\documentstyle[11pt,newpasp,twoside,epsf]{article}

\def\edcomment#1{\iffalse\marginpar{\raggedright\sl#1\/}\else\relax\fi}
\reversemarginpar

\begin{document}

\title{Stellar Collisions and Blue Straggler Formation}
 \author{James C.\ Lombardi, Jr.}
\affil{Vassar College, 124 Raymond Ave., Poughkeepsie, NY
12604-0562}
\author{Frederic A.\ Rasio}
\affil{MIT 6-201, Cambridge, MA 02139}

\begin{abstract}
We review recent 3D hydrodynamic calculations of stellar collisions
using the Smoothed Particle Hydrodynamics (SPH) method,
and we discuss the implications of the results for the formation
and evolution of blue stragglers in globular clusters.
We also discuss the construction of simple analytic models
for merger remnants,
approximating the mass loss, shock heating, mixing, and angular
momentum transport during a collision with simple algorithms
that can be calibrated using our 3D
hydrodynamic results.  The thermodynamic and chemical composition
profiles predicted by these simple models are compared with those
from our SPH simulations, demonstrating that our new models provide
accurate representations of true collisional merger remnants.
\end{abstract}

\section{Introduction}

\subsection{Stellar Interactions in Globular Clusters}

In the cores of dense globular clusters, the stellar collision time can
become
comparable to the cluster lifetime, implying that
essentially all stars must be affected by collisions (Hills \&
Day 1976). Stellar collisions have important consequences for the
overall dynamical evolution of dense star clusters. The
dissipation of kinetic energy in collisions and tidal encounters
and the formation of more massive objects through mergers tend to
accelerate core collapse. On the other hand, mass loss from
evolving collision products can provide indirect heating of the
cluster core, thereby slowing down core collapse.

In addition, in globular clusters, stellar collisions are thought
to produce a variety of interesting observable objects at rates
far exceeding those in the rest of the Galaxy. Low-mass X-ray
binaries and recycled pulsars are thought to be formed through
close dynamical interactions involving main-sequence or red-giant
stars and neutron stars (see, e.g., Davies et al.\ 1992; Di
Stefano \& Rappaport 1992; Rasio \& Shapiro 1991; Rasio et al.\
2000). Cataclysmic variables are also expected to be produced
through interactions involving white dwarfs (e.g., Di Stefano \&
Rappaport 1994; Davies 1997). Both white dwarfs and cataclysmic
variables have now been detected directly in many clusters
(Bailyn et al.\ 1996; Cool et al.\ 1998; Cool, in this volume;
Edmonds et al.\ 1999; Richer et al.\ 1997).

Interactions involving main-sequence (MS) stars are thought to be
related to the {\it blue straggler\/} phenomenon. Blue stragglers
are objects that appear as MS stars above the turnoff point in
the color-magnitude diagram (CMD) of a star cluster. All
observations suggest that they are indeed massive MS stars formed
through the merger of two (or more) lower-mass stars. In
particular, Saffer et al.\ (1997) and Sepinsky et al.\ (2000)
have directly measured the masses of several blue stragglers in
the cores of 47 Tuc and NGC 6397 and confirmed that they are well
above the MS turnoff mass (see also the article by Saffer in this
volume). Gilliland et al.\ (1998) have
demonstrated that the masses estimated from the pulsation
frequencies of four oscillating blue stragglers in 47 Tuc are
consistent with their positions in the CMD.

Mergers of MS stars can occur in at least two different ways:
following the physical collision of two single stars, or through
the coalescence of the two components in a close binary system
(Bailyn 1995; Leonard 1989; Livio 1993). Direct evidence for
binary progenitors has been found in the form of contact (W~UMa
type) binaries among blue stragglers in low-density globular
clusters such as NGC 5466 (Mateo et al.\ 1990) and M71 (Yan \&
Mateo 1994), as well as in many open clusters (see Ahumada \&
Lapasset 1995 for a recent survey).
At the same time, strong indication for a collisional origin
comes from detections by HST of large numbers of blue stragglers
concentrated in the cores of some of the densest clusters, such
as M15 (De Marchi \& Paresce 1994; Guhathakurta et al.\ 1996),
M30 (Yanny et al.\ 1994; Guhathakurta et al.\ 1998), NGC 6624
(Sosin \& King 1995), and M80 (Ferraro et al.\ 1999).

Collisions can happen directly between two single stars only in
the cores of the densest clusters, but, even in moderately dense
clusters, they can also happen indirectly, during resonant
interactions involving primordial binaries (Bacon et al.\ 1996;
Cheung et al.\ 2000; Davies \& Benz 1995; Sigurdsson et al.\
1994; Sigurdsson \& Phinney 1995). Observational evidence for the
existence of dynamically significant numbers of primordial binaries
in globular clusters is well
established (C{\^o}t{\'e} et al.\ 1994; Hut et al.\ 1992;
Rubenstein \& Bailyn 1997). Dynamical interactions between hard
primordial binaries and other single stars or binaries are
thought to be the primary mechanism for supporting a globular
cluster against core collapse (Gao et al.\ 1991; Goodman \& Hut 1989;
Rasio 2000). In addition, exchange interactions between primordial
binaries and compact objects can explain very naturally the
formation in globular cluster cores of large numbers of X-ray
binaries and recycled pulsars (see, e.g., Rasio, Pfahl, \&
Rappaport 2000 and references therein).

In all clusters, dynamical interactions between primordial
binaries can result in dramatically increased collision rates.
This is because the interactions are often resonant, with all
four stars involved remaining together in a small volume for a
long time ($\sim100-1000$ orbital times). For example, in the
case of an interaction between two typical hard binaries with
semi-major axes $\sim1\,$AU containing $\sim1\,M_\odot$ MS stars,
 the effective cross section for collision between
any two of the four stars involved is essentially equal to the
entire geometric cross section of the binaries (Bacon et al.\
1996; Cheung et al.\ 2000). This implies a collision rate
$\sim100$ times larger than what would be predicted for single
stars. Collisions involving more than two stars can be quite
common during binary--binary interactions, since the product of a
first collision between two stars expands adiabatically following
shock heating, and therefore has a larger cross section for
subsequent collisions with one of the other two stars
(Cheung et al.\ 2000).
These multiple collisions provide a natural explanation for the existence
 of blue stragglers that may have masses above {\it twice\/} the turnoff mass
(Saffer, in this volume; Sepinsky et al.\ 2000).

\subsection{Hydrodynamic Calculations of Stellar Mergers}

Benz \& Hills (1987; 1992) performed the first 3D
calculations of direct collisions between two MS stars, using the
Smoothed Particle Hydrodynamics (SPH) method.  An
important conclusion of their pioneering study was that stellar
collisions could lead to thorough mixing of the fluid. In
particular, they pointed out that the mixing of fresh hydrogen
fuel into the core of the merger remnant could reset the nuclear
clock of a blue straggler, allowing it to remain visible for a
full MS lifetime $\ga10^9\,$yr after its formation. In subsequent
work it was generally assumed that the merger remnants resulting
from stellar collisions were nearly homogeneous. Blue stragglers
would then start their life close to the zero-age MS, although
with an anomalously high helium abundance coming from the
hydrogen burning in the
 parent stars.
In contrast, little hydrodynamic mixing was expected to occur
during the much gentler process of binary coalescence, which
could take place on a stellar evolution timescale rather than on
a dynamical timescale (Bailyn 1992; Mateo et al.\ 1990).

On the basis of these ideas, Bailyn (1992) suggested a way of
distinguishing observationally between the two possible formation
processes. Blue stragglers made from collisions would have a
higher helium abundance in their outer layers than those made
from binary mergers, and this would generally make them appear
somewhat brighter and bluer. A detailed analysis was carried out
by Bailyn \& Pinsonneault (1995), who performed the first stellar
evolution calculations for blue stragglers assuming various
initial chemical composition profiles. To represent the
collisional case, chemically homogeneous initial profiles with
enhanced helium abundances were assumed. For the dense cluster
47~Tuc, the observed luminosity function and numbers of blue
stragglers appeared to be consistent with a collisional origin.

Lombardi, Rasio, \& Shapiro (1995, 1996) later re-examined the
question of hydrodynamic mixing during stellar collisions, using
higher-resolution SPH calculations. They found that, in contrast
to previous results, hydrodynamic mixing during collisions may be
very inefficient, and typical merger remnants produced by
collisions could be far from chemically homogeneous. In the case
of collisions between two nearly identical stars, the final
chemical composition profile is in fact very close to the initial profile
of the parent stars (Lombardi, Rasio, \& Shapiro 1995; Rasio
1996a). For two turnoff stars, this means that the core of the
merger remnant is mostly helium. Prompted by these results,
Sills, Bailyn \& Demarque (1995) started investigating the
possible consequences of blue stragglers being born unmixed.
Using stellar structure calculations, they compared the predicted
colors and luminosities of initially unmixed models with
observations of the six very bright blue stragglers in the core
of NGC 6397. They concluded that some of these blue stragglers
have observed colors that cannot be explained using unmixed
initial models (cf.\ Ouellette \& Pritchet 1998). Initially
homogeneous models, however, could reproduce all the
observations. In addition, unmixed models have very short MS
lifetimes and may be generally incompatible with the observed
numbers of blue stragglers in dense cluster cores.
The key to resolving this apparent inconsistency may be stellar
rotation, which can both provide more mixing and increase the
stellar evolution lifetime of collision products
(Sills et al.\ 2000; Sills, in this volume).

Our most recent hydrodynamic calculations (Sills \& Lombardi 1997;
Sills et al.\ 2000) improve on previous
studies (Benz \& Hills 1987, 1992; Lombardi et al.\ 1995, 1996;
Sandquist et al.\ 1997) by adopting more realistic initial
stellar models and by performing numerical calculations with
increased spatial resolution. The latest version of our SPH code
has been fully parallelized and
allows calculations to be performed with $N\sim 10^5-10^6$ SPH
particles, compared to the $N\sim10^3-10^4$ used in previous
studies. All spurious transport processes affecting SPH
calculations can be reduced significantly by increasing the
number of SPH particles (Lombardi et al.\ 1999). For the first
time in a study of stellar collisions, accurate initial stellar
models have been calculated using a state-of-the-art stellar
evolution code (the YREC code developed at Yale by Demarque and
collaborators; see Guenther et al.\ 1992 and Sills et al.\ 2000).
This allows us, in particular, to study chemical mixing and
stellar evolution of collision products in a completely
self-consistent way. The importance of using realistic initial
models (rather than simple polytropes, which were used in all
previous studies) in stellar collision
calculations was demonstrated by Sills \& Lombardi (1997). In
some cases, results based on calculations for polytropes can
actually lead to {\it qualitatively incorrect\/} conclusions
about the structure of the merger remnant.

The vast majority of 3D numerical studies of stellar collisions
have used the SPH method (see, e.g., Monaghan 1992 and Rasio \&
Lombardi 1999 for recent reviews). Indeed,
because of its Lagrangian nature, SPH presents some clear
advantages over more traditional grid-based methods for
calculations of stellar interactions. Most importantly, fluid
advection, even for stars with a sharply defined surface, is
accomplished without difficulty in SPH, since the particles
simply follow their trajectories in the flow. In contrast, to
track accurately the orbital motion of two stars across a large
3D grid can be quite tricky, and the stellar surfaces then
require a special treatment (to avoid ``bleeding''). A Lagrangian
scheme is also ideal for following hydrodynamic mixing, since the
chemical composition attached to each particle is simply advected
with that particle. This is a key advantage for the problems
discussed here, where the chemical composition profiles of merger
remnants must be determined accurately. SPH is also very
computationally efficient, since it concentrates the numerical
elements (particles) where the fluid is at all times, not wasting
any resources on emty regions of space. For this reason, with
given computational resources, SPH can provide higher averaged
spatial resolution than grid-based calculations, although
Godunov-type schemes such as PPM typically provide better
resolution of shock fronts. SPH also makes it easy to track the
hydrodynamic ejection of matter to large distances from the
central dense regions. This is a key advantage for calculating
processes such as collisions,
where ejection of significant amounts of matter through
rotational instabilities or gas shocking plays an important role.
Sophisticated nested-grid algorithms and more expensive
calculations are necessary to accomplish the same with grid-based
methods.

Spurious transport by SPH particles can render completely
unphysical the results of low-resolution hydrodynamics calculations.
For example, particle shear viscosity can lead to the
artificial generation of vorticity.
The strong differential rotation observed in collision products
 could also be damped rapidly, leading to spurious
transport of angular momentum away from the central region. In
addition, numerical viscosity can also lead to spurious entropy
production and this contributes systematic errors in all
thermodynamic properties of the fluid. When studying chemical
mixing during mergers, it is particularly important to calibrate
how much of the observed mixing is real, and how much results
from spurious diffusion of SPH particles. Based on a very
extensive study with test calculations (Lombardi et al.\ 1999),
we are confident that, with $\sim10^5-10^6$ SPH particles, and
with the latest improvements to our code
(see Lombardi et al.\ 1999
for details), spurious transport processes can be brought down to
an acceptable level.

\subsection{Thermal Relaxation and Stellar Evolution of Merger Remnants}

Hydrodynamic calculations are of course limited to
following the evolution of mergers on a dynamical timescale
($t_{dyn}\sim\,$hours for MS stars) but are not capable of
following processes taking place on a thermal timescale
($t_{th}\sim10^7\,$yr for MS stars). The final configurations
obtained at the end of hydrodynamic calculations are very close
to {\it hydrostatic equilibrium\/}, but are generally far from
{\it thermal equilibrium\/}. Therefore, the merger remnant will
need to contract (on its Kelvin time $t_{th}$) before it can be
treated as a star again. This thermal relaxation of an object
initially far from equilibrium cannot be calculated with most
standard stellar evolution codes, which assume small departures
from thermal equilibrium. However, to compute theoretically the
subsequent stellar evolution of the object, and to make
comparisons with observations, it is necessary to perform
detailed calculations of this thermal relaxation phase, taking
into account processes that are not always incorporated into
existing stellar evolution codes. As the object evolves towards
thermal equilibrium, many processes can lead to additional mixing
of the fluid. In particular, thermal convection, which is well
known to occur during the evolution of ordinary pre-MS stars,
could produce significant mixing in blue stragglers (see Leonard
\& Livio 1995). Even if ordinary convection does not occur, local
thermal instabilities such as semi-convection and
thermohaline-type instabilities (Spruit 1992) can develop,
leading to mixing on the local radiative damping timescale.
Indeed, some of our previous hydrodynamic calculations (Lombardi
et al.\ 1996) suggest that many merger remnants, although
convectively stable, have temperature and chemical composition
profiles with large thermally unstable regions. For the rapidly
(and differentially) rotating products of grazing collisions,
meridional circulation, as well as various rotationally-induced
instabilities (see, e.g., Tassoul 1978), can lead not only to
mixing, but also to angular momentum transport and mass loss.
Loss of angular momentum through magnetic braking or coupling of
the star to an outer disk of ejected material could also play an
important role in spinning down the merger remnant.

For each final configuration obtained with our SPH code, we
calculate the thermal relaxation of the remnant using a modified
version of the Yale stellar evolution code  (YREC, see Sills at al.\ 2000
and Sills, in this volume), which follows the evolution of a rotating star
both on a thermal timescale and on a nuclear timescale (Sills
1998). YREC evolves a star through a sequence of models of
increasing age, solving the linearized stellar evolution
equations for interior profiles such as chemical composition,
pressure, temperature, density and luminosity.  All relevant
nuclear reactions (including pp-chains, the CNO cycle,
triple-$\alpha$ reactions and light element reactions) are
treated. Recent opacity tables are used (ensuring that the
remnant's position in a CMD can be accurately determined).  YREC
incorporates standard treatments of convection, semi-convection
and rotationally-induced mixing and angular momentum transport
through processes such as dynamical and secular shear
instabilities, meridional circulation, and the
Goldreich-Schubert-Fricke instability (see Guenther et al.\ 1992
for details). The free parameters in the code (the mixing length
and parameters that set the efficiency of angular momentum
transport and rotational chemical mixing) are set by calibrating
a solar mass and solar metallicity model to the Sun.
For blue stragglers, the various transport processes can
potentially carry fresh hydrogen fuel into the stellar core and
thereby extend the MS lifetime of the remnant.  Furthermore, any
helium mixed into the outer layers affect the opacity and
hence the remnant's position in a CMD.

This treatment, combining hydrodynamics and stellar evolution,
attempts to provide a
complete, self-consistent description of blue straggler formation
and evolution, assuming a collisional origin. Our first results
demonstrating the feasibility of this basic approach have
been presented in Sills et al.\ (2000; see also Sills, in this volume).
These results,
although encouraging, have revealed a number of important
theoretical problems that will have to be addressed carefully in
future work. Most importantly, we find that even slightly
off-axis collisions produce merger remnants that can be rotating
near break-up. As they contract to the MS, these objects must
lose some mass and a large amount of angular momentum. The most
likely angular momentum loss mechanism, which we plan to study in
detail, is magnetic coupling to an outer disk (containing the
material ejected during the initial collision). Ordinary magnetic
braking (as in pre-MS stars) is not likely to be important since,
at least based on the calculations we performed so far, typical
merger remnants never develop outer convective envelopes.

\section{SPH Calculations of Stellar Collisions}

\subsection{Implementation of the SPH Code}

Our most recent hydrodynamic calculations have been done using a
new, parallel version of the SPH code developed originally by
Rasio (1991) specifically for the study of hydrodynamic stellar
interactions (see, e.g., Faber \& Rasio 2000; Lai et al.\ 1993;
Lombardi et al.\ 1995, 1996; Rasio \& Livio 1996; Rasio \&
Lombardi 1999; Rasio \& Shapiro 1991, 1992, 1995).

The gravitational
field in our code is calculated on a 3D grid using an FFT-based
convolution algorithm. The density field is placed on a 3D grid by a
cloud-in-cell method, and convolved with a kernel function
calculated once during the initialization of the simulation.  We
use zero-padding of our grids to obtain correct boundary
conditions for an isolated system, at the expense of memory
storage.  Gravitational forces are calculated from the
gravitational potential by finite differencing on the grid, and
then interpolated for each particle using the same cloud-in-cell
assignment. For most problems, the CPU time is
dominated by the calculation of the gravitational field. The
current implementation is based on the high-performance FFTW
parallel routines (Frigo \& Johnson 1997). Our SPH code was
parallelized using MPI (the
message passing interface), to run efficiently on multiple
processors. In benchmarking tests, we have found that our
parallel code scales very well when using up to $\sim32$
processors on a distributed shared-memory supercomputer. Our
parallel code now allows us to perform SPH calculations using up
to $\sim10^6$ particles, which provides a spatial resolution
$\sim0.01\,R$ for colliding stars of radius $R$. For more details
on the current implementation of our parallel code, see Faber \&
Rasio (2000).

Local densities and hydrodynamic forces at each particle position
are calculated by smoothing over $N_N$ nearest neighbors.  The
size of each particle's smoothing kernel is evolved in time to
keep $N_N$ close to a predetermined optimal value.  For the
high-resolution calculations in this paper the optimal number of
neighbors was set at $N_N = 100$.  Neighbor lists for each
particle are recomputed at every iteration using a linked-list,
grid-based parallel algorithm.  Note that neighborhood is not a
symmetric property, as one particle can be included in another's
neighbor list but not vice versa.  Pressure and artificial
viscosity forces are calculated by a ``gather-scatter''
method, looping over all particles in turn, gathering the force
contribution on each particle from its neighbors, and scattering
out the equal and opposite force contribution on the neighbors
from each particle.

A number of variables are associated with each particle $i$,
including its mass $m_i$, position ${\bf r}_i$, velocity ${\bf
v}_i$, entropic variable $A_i$ and numerical smoothing length
$h_i$.  The entropic variable $A$ is a measure of the fluid's
compressibility and is closely related (but not equal) to
specific entropy: for example, both $A$ and specific entropy are
conserved in the absence of shocks. For convenience, we refer to
$A$ simply as ``entropy.''  In this paper we adopt an equation of
state appropriate for a monatomic ideal gas: the adiabatic index
$\Gamma=5/3$ and $P_i=A_i\rho_i^{\Gamma}$, where $P_i$ and
$\rho_i$ are the density and pressure of particle $i$.

The equation of motion for the SPH particles can be written as
\begin{equation}
           {d {\bf v}_i\over d t} = {\bf a}^{(Grav)}_i+{\bf
           a}^{(SPH)}_i,
\end{equation}
where ${\bf a}^{(Grav)}_i$ is the gravitational acceleration and
\begin{equation}
{\bf a}^{(SPH)}_i=-\sum_j m_j \left[\left({p_i\over\rho_i^2}+
    {p_j\over\rho_j^2}\right)+\Pi_{ij}\right]{\bf \nabla}_i W_{ij}.
    \label{fsph}
\end{equation}
The summation in equation (\ref{fsph}) is over neighbors and
$W_{ij}$ is a symmetrized smoothing kernel (see Rasio \& Lombardi
1999 for details).  The artificial viscosity (AV) term $\Pi_{ij}$
ensures that correct jump conditions are satisfied across
(smoothed) shock fronts, while the rest of equation~(\ref{fsph})
represents one of many possible SPH-estimators for the
acceleration due to the local pressure gradient (see, e.g.,
Monaghan 1985).

The rate of increase, due to shocks, for the entropy $A_i$ of
particle $i$ is given by
\begin{equation}
{dA_i\over dt}={\Gamma-1\over 2\rho_i^{\Gamma-1}}\,
     \sum_jm_j\,\Pi_{ij}\,\,({\bf v}_i-{\bf v}_j)\cdot{\bf \nabla}_i
     W_{ij}.
        \label{adot}
\end{equation}
We adopt the form of AV proposed
by Balsara (1995):
\begin{equation}
\Pi_{ij}= \left({p_i\over\rho_i^2}+{p_j\over\rho_j^2}\right)
        \left(-\alpha \mu_{ij} + \beta \mu_{ij}^2\right),
        \label{piDB}
\end{equation}
where
\begin{equation}
\mu_{ij}=\cases{ {({\bf v}_i-{\bf v}_j)\cdot({\bf r}_i-{\bf
r}_j)\over h_{ij}(|{\bf r}_i -{\bf
r}_j|^2/h_{ij}^2+\eta^2)}{f_i+f_j \over 2 c_{ij}}& if $({\bf
v}_i-{\bf v}_j)\cdot({\bf r}_i-{\bf r}_j)<0$\cr
             0& if $({\bf v}_i-{\bf v}_j)\cdot({\bf r}_i-{\bf
             r}_j)\ge0$\cr}.
                \label{muDB}
\end{equation}
The terms $h_{ij}$ and $c_{ij}$ are, respectively, the average
smoothing length and sound speed associated with particles $i$ and
$j$.  Here $f_i$ is the so-called form function for particle $i$,
defined by
\begin{equation}
f_i={|{\bf \nabla}\cdot {\bf v}|_i \over |{\bf \nabla}\cdot {\bf
v}|_i +|{\bf \nabla}\times {\bf v}|_i + \eta' c_i/h_i, }
\label{fi}
\end{equation}
where
\begin{equation}
({\bf \nabla}\cdot {\bf v})_i={1 \over \rho_i}\sum_j m_j
        ({\bf v}_j-{\bf v}_i)\cdot{\bf \nabla}_i W_{ij}. \label{divv}
\end{equation}
and
\begin{equation}
({\bf \nabla}\times {\bf v})_i={1 \over \rho_i}\sum_j m_j
        ({\bf v}_i-{\bf v}_j)\times{\bf \nabla}_i W_{ij}. \label{curlv}
\end{equation}
The function $f_i$ acts as a switch, approaching unity in regions
of strong compression ($|{\bf \nabla}\cdot {\bf v}|_i >>|{\bf
\nabla}\times {\bf v}|_i$) and vanishing in regions of large
vorticity ($|{\bf \nabla}\times {\bf v}|_i >>|{\bf \nabla}\cdot
{\bf v}|_i$). Consequently, this AV has the advantage that it is
suppressed in shear layers.  In this paper we use $\eta^2=0.01$,
$\eta'=10^{-5}$ and $\alpha=\beta=\Gamma/2$.  This choice of AV
treats shocks well, while introducing only relatively small
amounts of numerical viscosity (Lombardi et al.\ 1999).

For stability, the timestep must satisfy a modified Courant
condition, with $h_i$ replacing the usual grid separation. For
accuracy, the timestep must be a small enough fraction of the
dynamical time.  To determine the timestep, we follow the
prescription proposed by Monaghan (1989), which allows for an
efficient use of computational resources.  This method sets
\begin{equation}
\Delta t=C_N\,{\rm Min}(\Delta t_1,\Delta t_2), \label{dt}
\end{equation}
where the constant dimensionless Courant number $C_N$ typically
satisfies $0.1\la C_N \la 0.8$, and where
\begin{eqnarray}
\Delta t_1 &=&{\rm Min}_i\,(h_i/\dot v_i)^{1/2}, \label{dt1} \\
\Delta t_2 &=&{\rm Min}_i\left( {h_i \over c_i+k\left(\alpha
c_i+\beta {\rm Max_j}|\mu_{ij}|\right)} \right),  \label{good.dt}
\end{eqnarray}
with $k$ being a constant of order unity.

Our simulations employ equal-mass SPH particles, both to keep the
resolution high in the stellar cores and to minimize spurious
mixing during the simulation (see Lombardi et al.~1999 for a
discussion of spurious transport induced by unequal mass
particles).  Due to extreme central densities in the parent stars
(especially turnoff stars), the Courant stability condition
requires exceedingly small timesteps. Consequently, an elapsed
physical time of one hour in our high-resolution calculations can
require roughly 1000 iterations and 100 CPU hours on an SGI/Cray
Origin2000 supercomputer.

\subsection{Initial Conditions}

All our recent simulations use parent star models calculated with
YREC, as discussed in Sills \& Lombardi (1997) and Sills et al.\
(2000). In particular, we have evolved (non-rotating) MS stars of
total mass $M=0.6$ and $0.8\,M_\odot$ with a primordial helium
abundance $Y=0.25$ and metallicity $Z=0.001$ for 15 Gyr, the
amount of time needed to exhaust the hydrogen at the center of
the $0.8\,M_\odot$ star (a typical globular cluster turnoff mass).
The total helium mass fractions for the $0.6$ and $0.8\,M_\odot$
parent stars are 0.286 and 0.395, and their radii are
$0.517\,R_\odot$ and $0.955\,R_\odot$, respectively.  From the
pressure and density profiles of these models, we compute the
entropy profile and assign values of $A$ to SPH particles
accordingly.  In addition, the chemical abundance profiles of 15
different elements are used to set the composition of the SPH
particles. To minimize numerical noise, each parent star's SPH
model is relaxed to equilibrium using an artificial drag force on
the particles, and then these relaxed models are used to initiate
the collision calculations.

The stars are initially non-rotating and separated by $5\,R_{TO}$,
where $R_{TO}=0.955\,R_\odot$ is the radius of the turnoff
($M=0.8\,M_\odot$) star.  The initial velocities are calculated by
approximating the stars as point masses on an orbit with zero
orbital energy and a pericenter separation $r_p$.  A Cartesian
coordinate system is chosen such that these hypothetical point
masses of mass $M_1$ and $M_2$ would reach pericenter at positions
$x_i=(-1)^{i}(1-M_i/(M_1+M_2))r_p$, $y_i=z_i=0$, where $i=1,2$ and
$i=1$ refers to the more massive star.  The orbital plane is
chosen to be $z=0$.  With these choices, the center of mass
resides at the the origin.

\subsection{Sample of Numerical Results}

Table~\ref{tbl-summary} summarizes the parameters and results of
the simulations presented in this section. The first column gives
the name by which the calculation is referred to in this paper;
we use lower-case letters to distinguish the present calculations
from the corresponding calculations involving polytropic parent
stars presented in Lombardi et al.\ (1996). The second and third
columns give the masses $M_1$ and $M_2$ of the parent stars.
Column~(4) gives the ratio $r_p/(R_1 +R_2)$, where $r_p$ is the
pericenter separation for the initial orbit and $R_1+R_2$ is the
sum of the two (unperturbed) stellar radii. This ratio has the
value $0$ for a head-on collision, and $\sim 1$ for a grazing
encounter. Column~(5) gives the number of SPH particles.
Column~(6) gives the final time $t_f$ at which the calculation
was terminated. Column~(7) gives the total angular momentum of
the merger remnant.  Column~(8) gives the ratio $T/|W|$ of
rotational kinetic energy to gravitational binding energy of the
(bound) merger remnant in its center-of-mass frame at time $t_f$.
Since the amount of mass ejected during a parabolic collision is
quite small, the merger remnant never acquires a large recoil
velocity.  The Case~e, f and k simulations implemented $N_N=100$
neighbors per particle on average, while Case~j$^\prime$
implemented an average of 32 neighbors per particle.  Case
j$^\prime$ is a low-resolution calculation of a nearly head-on
collision used in Sills et al.\ (2000) to help demonstrate the
applicability of our approach for converting the final SPH
remnant model to a YREC starting model.

\begin{table}
\caption{Off-axis Collisions of Realistically Modeled Parent Stars
\label{tbl-summary}}
\begin{tabular}{cccccccc}
\\
Case & $M_1$   & $M_2$   & $r_p$ & $N$ & $t_f$ & $J$ & $T/|W|$ \\
& $[M_\odot]$   & $[M_\odot]$   & $[R_1+R_2]$ &  & [hours] & [g
cm$^2$ s$^{-1}$] &  \\
(1)&(2)&(3)&(4)&(5)&(6)&(7)&(8) \\
\tableline e & 0.8 & 0.6 & 0.25 & $1.05\times 10^5$ & 11.10 &
$2.0\times
10^{51}$ & 0.101 \\
f & 0.8 & 0.6 & 0.50 & $1.05\times 10^5$ & 24.64 & $2.8\times
10^{51}$ & 0.119 \\
k & 0.6 & 0.6 & 0.25 & $9\times 10^4$ & 11.10 & $1.3\times
10^{51}$ & 0.085 \\
j$^\prime$ & 0.6 & 0.6 & 0.01 & $1.6\times 10^3$ & 9.25 &
$2.1\times 10^{50}$ & 0.005 \\
\end{tabular}
\end{table}

\begin{figure}
\plotone{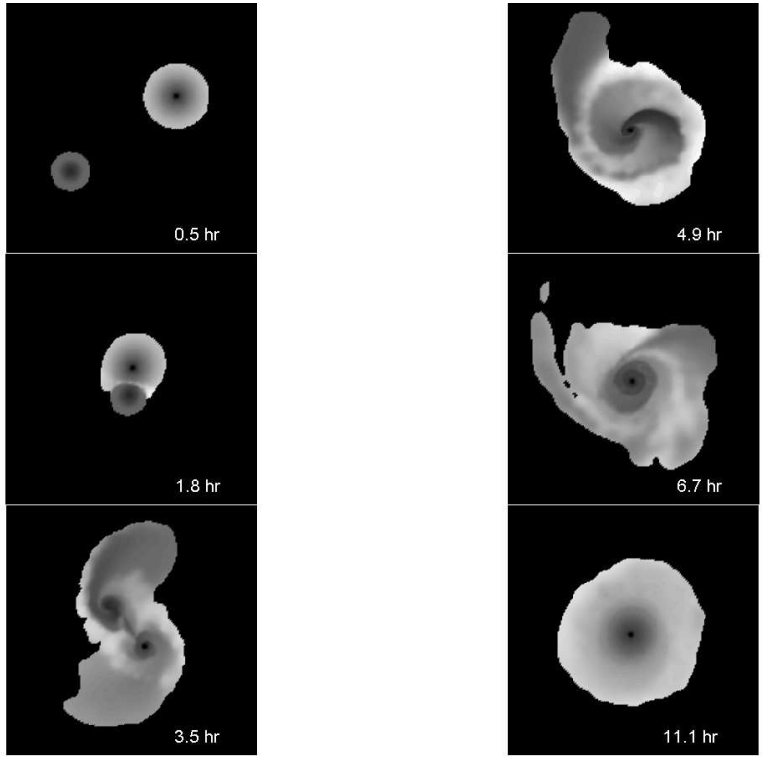} \caption {Snapshots of the entropy $A$
in the orbital plane for case~e, a parabolic collision between
parent stars of masses $M_1=0.8M_\odot$ and $M_2=0.6M_\odot$ at a
pericenter separation $r_p=0.25(R_1+R_2)$.  Fluid with
$\log_{10}A \leq 13.2$ (cgs) is represented with black.  For
larger specific entropies, the fluid is represented with lighter
shades of gray until white is reached at $\log_{10}A=16.1$ (cgs).
Only the fluid at densities exceeding $6.8\times 10^{-3}$
g/cm$^3$ is shown. Each frame spans an area $7.6R_\odot\times
7.6R_\odot$, with the center of mass at the center of the frame.
The time elapsed since the initial conditions of the calculation
is shown in the lower right corner of each frame. \label{Afig} }
\end{figure}

Figure \ref{Afig} illustrates the dynamical evolution of the
entropy $A$ in the orbital plane for case~e: a terminal age
main-sequence star ($M_1=M_{TO}=0.8M_\odot$) collides with a
slightly less massive star ($M_2=0.75\,M_{TO}=0.6M_\odot$).  The
parabolic trajectory has a pericenter separation $r_p=0.25
(R_1+R_2)$.  The first collision occurs $\sim 1.8$ hours after
the initial configuration of the calculation. The impact disrupts
the outer layers of the two parents, but leaves their inner cores
essentially undisturbed.  The two components withdraw to
apocenter at $t\simeq 3.2$ hours, with a bridge of low entropy
fluid from the smaller parent connecting the cores.  As the cores
fall back towards one another and merge, a complicated shock
heating pattern ensues, with the strongest shocks occurring near
the leading edge of the fluid from the smaller parent.  The cores
themselves are largely shielded from the shock heating, and the
lowest entropy fluid settles down to the center as the remnant
approaches dynamical equilibrium.  The merger remnant is rapidly
and differentially rotating, and shear makes the bound fluid
quickly approach axisymmetry within a few dynamical timescales.
Visualizations of our collision calculations can be found at
http://research.amnh.org/$\sim$summers/scviz/lombardi.html.

\begin{figure}
\plotone{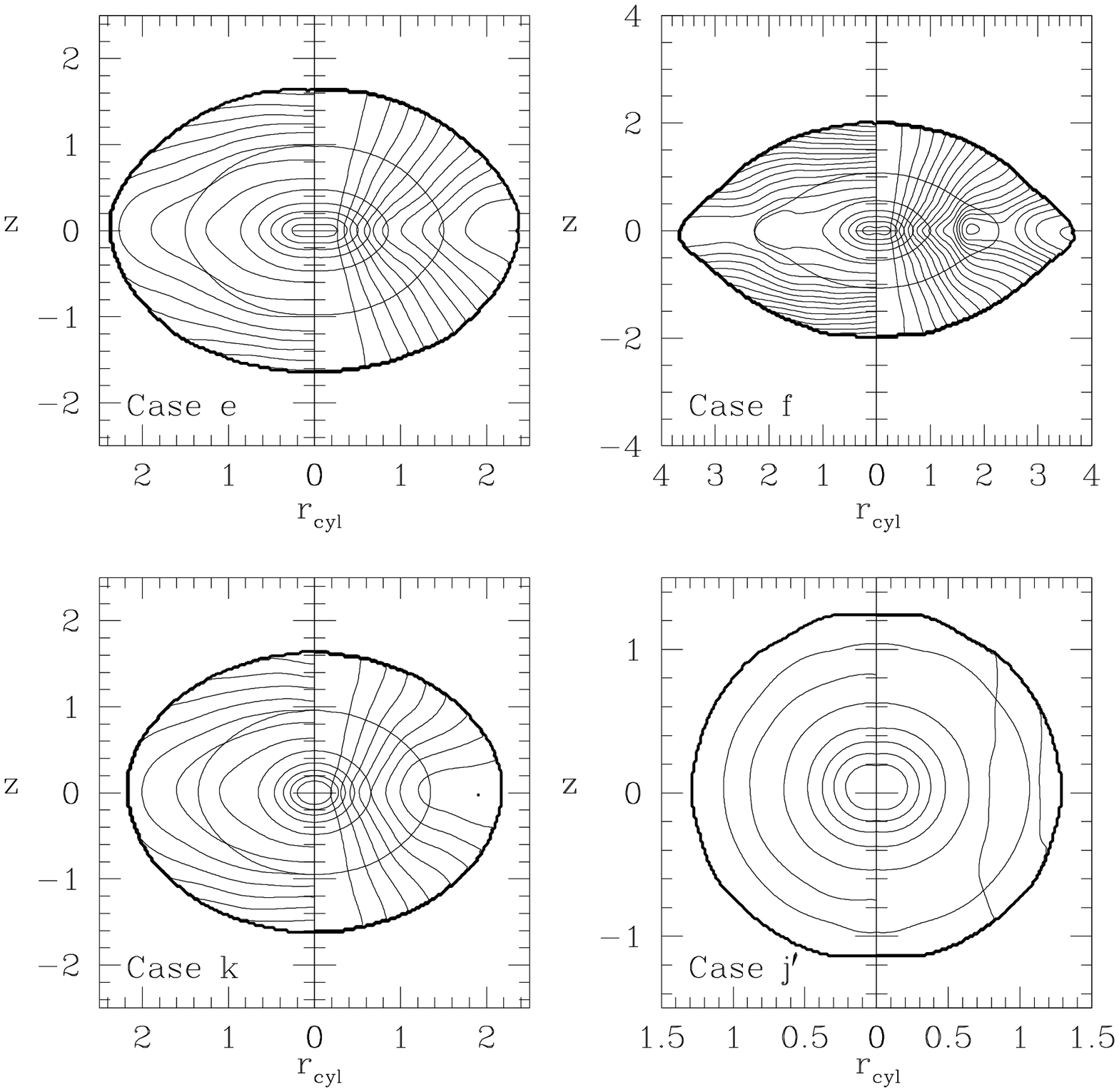} \caption {Constant density, specific
angular momentum and entropy contours at the end of each
simulation in slices containing the remnants' rotation axes. Here
$r_{cyl}$ is the cylindrical radius (measured from the rotation
axis) in units of the turnoff radius ($0.96 R_\odot$). The closed
loops that extend to both the right and left halves of each plot
correspond to the isodensity surfaces enclosing 15\%, 30\%, 45\%,
60\%, 75\% and 90\% of the remnant mass, while the thick outermost
bounding curve encloses 95\% of the total mass.  The left half of
each plot shows constant entropy contours, with a linear spacing
of $1.04\times 10^{15}$ (cgs), while the right half of each plot
shows constant specific angular momentum contours, with a linear
spacing of $2.66\times 10^{17}$ cm$^2$ s$^{-1}$. \label{all} }
\end{figure}

Figure \ref{all} shows contours of density $\rho$, entropy $A$ and
$z-$component of the specific angular momentum $j$ for each
collision remnant in a slice containing its rotation axis.  The
contours shown represent an average over the azimuthal angle, as
the remnant approaches axisymmetry at the end of the simulation.
The conditions for dynamical stability against axisymmetric
perturbations are that (1) the entropy $A$ never decreases
outward, and (2) on each constant $A$ surface, the specific
angular momentum $j$ increases from the poles to the equator (the
Hoiland criterion; see, e.g., Tassoul 1978, Chap.~7).  Both of
these conditions are met throughout the preponderance of mass in
our remnant models.

Figure \ref{all} also reveals that for our rapidly rotating
remnants (cases e, f and k) the specific angular momentum $j$
(and hence the angular velocity $\omega$, related by
$j=r_{cyl}\omega^2$) varies significantly on the cylindrical
surfaces of constant $r_{cyl}$. These rotating remnants are
therefore not barotropes, by definition.  For a rotating,
chemically homogeneous star, stable thermal equilibrium requires
$\partial\omega/\partial z=0$, where $z$ is measured along the
direction of the rotation axis (the Goldreich-Schubert criterion;
see, e.g., Tassoul 1978, Chap.~7).  In chemically inhomogeneous
stars, regions with a sufficiently large and stabilizing
composition gradient can in principle still be thermally stable
even with $\partial\omega/\partial z \neq 0$. However, all of our
remnants, rotating or not, are far from thermal equilibrium.

\begin{figure}
\plotone{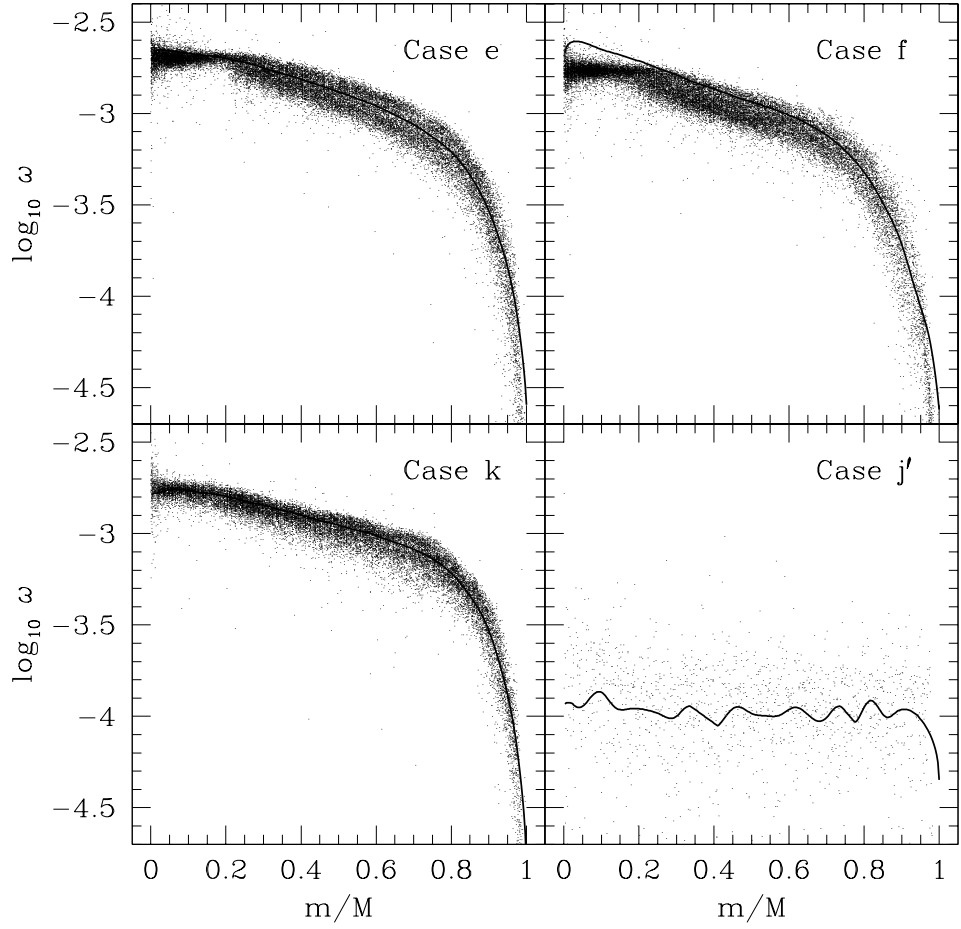} \caption {Angular velocity
$\omega$ as a function of the enclosed mass fraction $m/M$ in the
final merger remnants.  The points represent particle values from
an SPH simulation, with only half of the particles displayed in
cases e, f and k.  The solid lines show the $\omega$ profiles of
the initial YREC models, generated by a procedure that constrains
the entropy and specific angular momentum profiles of each model
to be that given by the SPH results.  Units of $\omega$ are
rad$\,{\rm s}^{-1}$. \label{alllogomegam} }
\end{figure}

To convert our fully three-dimensional hydrodynamic results into
the one-dimensional YREC format, we first average the entropy and
specific angular momentum values in $\sim 25$ bins in enclosed
mass fraction.  With the $A$ and $j$ profiles given, the
structure of the rotating remnant is uniquely determined in the
formalism of Endal and Sofia (1976), by integrating the general
form of the equation of hydrostatic equilibrium [see their eq.\
(9)]. To do so, we implement an iterative procedure in which
initial guesses at the central pressure and angular velocity are
refined until a self-consistent YREC model is converged upon.
Although the pressure, density, and angular velocity profiles of
the resulting YREC model cannot also be simultaneously
constrained to equal the corresponding (averaged) SPH profiles,
the differences are slight (see, for example, the $\omega$
profiles in Figure \ref{alllogomegam}).  Indeed, the subsequent
evolutionary tracks are not significantly different for remnant
models that use the form of the $\omega$ profile taken directly
from SPH.

Examples of results from YREC integrations for collision products
and predictions for the location of rotating blue stragglers in
the CMD are presented in the article by Sills (in this volume).

\section{Simple Models for Merger Remnants \label{simple}}

The results of our hydrodynamic calculations can be used to
construct a set of simple analytic fitting formulae that can be
used in other studies to construct approximate models for stellar
collision products, without resorting to full hydrodynamic
calculations. In particular, these results can be used in studies
of dense star cluster dynamics that take into account the
important effects of stellar collisions on the overall dynamical
evolution of the cluster, and in scattering experiments for
binary--single and binary--binary interactions that take into
account the (very frequent) stellar collisions occurring during
resonant interactions (instead of the far more costly approach of
running a low-resolution SPH simulation each time a collision
takes place; cf.\ Davies et al.\ 1994).

We begin with two (non-rotating) parent star models, specifying
initial profiles for the stellar density $\rho$, pressure $P$,
and abundance of chemicals. The profile for the entropic variable
$A\equiv P/\rho^{\Gamma}$ can also be calculated easily and is of
central importance. Fluid elements with low values of $A$ sink to
the bottom of a gravitational potential well, and the $A$ profile
of a star in stable dynamical equilibrium increases radially
outwards. Indeed, it is easy to show that the condition $dA/dr
>0$ is equivalent to the usual Ledoux criterion for convective
stability of a nonrotating star (Lombardi et al.\ 1996). Since
the quantity $A$ depends only on the chemical composition and the
entropy, it remains constant in the absence of shocks. Our
algorithms are based upon closely modeling the results of SPH
calculations presented in Lombardi et al.\ (1996) (for collisions
of polytropic stars) as well as in Sills \& Lombardi (1997) and
\S2.3 of this paper (for collisions of more realistically modeled
stars). We model mass loss, shock heating, fluid mixing, and the
angular momentum distribution.

\subsection{Mass Loss} \label{massloss}

The velocity dispersion of globular cluster stars is typically
only $\sim 10$ km s$^{-1}$, which is much smaller than the escape
velocity from the surface of a MS star [for example, a star of
mass $M=0.8 M_\odot$ and radius $R=R_\odot$ has an escape velocity
$(2GM/R)^{1/2}=552$ km s$^{-1}$].  For this reason, collisional
trajectories are well approximated as parabolic, and the mergers
are relatively gentle: the mass lost is never more than about 8\%
of the total mass in the system (mass loss with hyperbolic
trajectories is treated by Lai, Rasio, \& Shapiro 1994).
Furthermore, most main sequence stars in globular clusters are
not rapidly rotating, and it is a good approximation to treat the
initial parent stars as non-rotating.

Given models for the parent stars (see Table \ref{tbl-1}), we
first determine the mass lost during a collision.  Inspection of
hydrodynamic results for collisions between realistically modeled
stars, as well as for collisions between polytropes, suggests
that the fraction of mass ejected can be estimated approximately
by
\begin{equation}
f_L = c_1 {\mu \over M_1+M_2} {R_1 +R_2 \over R_{1,0.7} +
R_{2,0.7} +c_2 r_p}, \label{fL}
\end{equation}
where $c_1$ and $c_2$ are dimensionless constants which we take
to be 0.1 and 3, $\mu\equiv M_1M_2/(M_1+M_2)$ is the reduced mass
of the parent stars, $R_i$ is the radius of parent star $i$,
$R_{i,0.7}$ is the radius of parent star $i$ at an enclosed mass
fraction $m/M_i=0.7$, and $r_p$ is the periastron separation for
the initial parabolic orbit.  While developing equation
(\ref{fL}) we searched for a relation that accounted for the {\it
mass distribution} (not just the total masses and radii) of the
parent stars in some simple way.  The more diffuse the outer
layers of the parents, the longer the stellar cores are able to
accelerate towards each other after the initial impact: the
$R_{1,0.7} + R_{2,0.7}$ in the denominator of equation (\ref{fL})
accounts for this increased effective collision speed for parents
whose mass distributions are centrally concentrated.  The
dependence on $\mu$ in equation (\ref{fL}) arises from the
expectation that the mass loss will be roughly proportional to
the kinetic energy at impact, and from the fact that a simple
rescaling of the stellar masses ($M_i \rightarrow k M_i$) in a
hydrodynamic simulation leaves $f_L$ unchanged. This method
yields remnant masses which are seldom more than $0.01M_\odot$
different than what is given by a hydrodynamic simulation (see
the last two columns of Table \ref{tbl-2}); this is clearly a
significant improvement over neglecting mass loss completely,
which sometimes can overestimate the remnant mass by more than
$0.1 M_\odot$.

We distribute the mass loss between the two parent stars in such
a way that the outermost fluid layers of each parent which are
retained in the remnant have the same entropic variable $A$.

\begin{table}
\caption{Parent Star Characteristics \label{tbl-1}}
\begin{tabular}{clll}
\\
Structure type & $M [M_\odot]$ & $R
[R_\odot]$   & $R_{0.7} [R_\odot]$ \\
\tableline
Polytropic & 0.8 & 0.955 & 0.38 \\
Polytropic & 0.6 & 0.54 & 0.36 \\
Polytropic & 0.4 & 0.35 & 0.25 \\
Polytropic & 0.16& 0.15 & 0.11 \\
Realistic & 0.8 & 0.955 & 0.284 \\
Realistic & 0.6 & 0.517 & 0.262 \\
Realistic & 0.4 & 0.357 & 0.227\\
\end{tabular}
\end{table}

\begin{table}
\caption{Mass loss \label{tbl-2}}

\begin{tabular}{cccccccc}
\\
Case\tablenotemark{a} & $M_1 [M_\odot]$ & $M_2 [M_\odot]$ &
$r_p/(R_1+R_2)$ & $f_{L,SPH}$\tablenotemark{b}  &
$f_L$\tablenotemark{c} & $M_{SPH} [M_\odot]$\tablenotemark{d} &
$M_r [M_\odot]$\tablenotemark{e}\\
\tableline
A & 0.8 & 0.8 & 0.00  & 0.064 & 0.063 & 1.50 & 1.50\\
B & 0.8 & 0.8 & 0.25  & 0.023 & 0.022 & 1.56 & 1.57\\
C & 0.8 & 0.8 & 0.50  & 0.012 & 0.013 & 1.58 & 1.58\\
D & 0.8 & 0.6 & 0.00  & 0.057 & 0.050 & 1.32 & 1.33\\
E & 0.8 & 0.6 & 0.25  & 0.024 & 0.020 & 1.37 & 1.37\\
F & 0.8 & 0.6 & 0.50  & 0.008 & 0.012 & 1.39 & 1.38\\
G & 0.8 & 0.4 & 0.00  & 0.056 & 0.046 & 1.13 & 1.14\\
H & 0.8 & 0.4 & 0.25  & 0.028 & 0.018 & 1.17 & 1.18\\
I & 0.8 & 0.4 & 0.50  & 0.008 & 0.011 & 1.19 & 1.19\\
J & 0.6 & 0.6 & 0.00  & 0.049 & 0.038 & 1.14 & 1.16\\
K & 0.6 & 0.6 & 0.25  & 0.028 & 0.018 & 1.17 & 1.18\\
L & 0.6 & 0.6 & 0.50  & 0.022 & 0.012 & 1.17 & 1.19\\
N & 0.6 & 0.4 & 0.25  & 0.029 & 0.017 & 0.97 & 0.98\\
O & 0.6 & 0.4 & 0.50  & 0.010 & 0.011 & 0.99 & 0.99\\
P & 0.4 & 0.4 & 0.00  & 0.037 & 0.035 & 0.77 & 0.77\\
Q & 0.4 & 0.4 & 0.25  & 0.029 & 0.017 & 0.78 & 0.79\\
R & 0.4 & 0.4 & 0.50  & 0.010 & 0.011 & 0.79 & 0.79\\
S & 0.4 & 0.4 & 0.75  & 0.008 & 0.008 & 0.79 & 0.79\\
T & 0.4 & 0.4 & 0.95  & 0.011 & 0.007 & 0.79 & 0.79\\
U & 0.8 & 0.16& 0.00  & 0.026 & 0.031 & 0.94 & 0.93\\
V & 0.8 & 0.16& 0.25  & 0.025 & 0.012 & 0.94 & 0.95\\
W & 0.8 & 0.16& 0.50  & 0.021 & 0.007 & 0.94 & 0.95\\
a & 0.8 & 0.8 & 0.00  & 0.080 & 0.084 & 1.47 & 1.47\\
e & 0.8 & 0.6 & 0.25  & 0.029 & 0.022 & 1.36 & 1.37\\
f & 0.8 & 0.6 & 0.50  & 0.014 & 0.013 & 1.38 & 1.38\\
g & 0.8 & 0.4 & 0.00  & 0.063 & 0.057 & 1.12 & 1.13\\
k & 0.6 & 0.6 & 0.25  & 0.032 & 0.020 & 1.16 & 1.18\\
\tablenotetext{a}{Capital letters refer to collisions of
polytropic stars; lower case letters refer to cases involving more
realistically modeled parent stars} \tablenotetext{b}{The
fractional mass loss as determined by an SPH simulation}
\tablenotetext{c}{The fractional mass loss as estimated by
equation (\ref{fL})} \tablenotetext{d}{The remnant mass as
determined by an SPH simulation} \tablenotetext{e}{The remnant
mass as estimated by $(1-f_L)(M_1+M_2)$}
\end{tabular}
\end{table}

\subsection{Shock Heating} \label{shock}

Shocks increase the value of the fluid's entropic variable
$A=P/\rho^{\Gamma}$ (see eq.~\ref{adot}). The distribution and
timing of shock heating during a collision involve numerous
complicated processes (see Fig.~\ref{Afig}): an initial shock
front is generated at the interface between the stars during
impact, the oscillating merger remnant sends out waves of shock
rings, and finally the outer layers of the remnant are shocked as
gravitationally bound ejecta fall back to the remnant surface.
Our goal is not to {\it derive} approximations describing the
shock heating during each of these stages, but rather empirically
to determine physically reasonable relations which fit the
available SPH data.

Let $A$ and $A_{init}$ be, respectively, the final and initial
values of the entropic variable for some particular fluid
element.  We used the results of hydrodynamic calculations to
examine how the change $A-A_{init}$, as well as the ratio
$A/A_{init}$, depended on a variety of functions of $P_{init}$
(the initial pressure) and $A_{init}$.  Our search for a simple
means of modeling this dependence was guided by a handful of
features evident from hydrodynamic simulations: (1) fluid deep
within the parents are shielded from the brunt of the shocks, (2)
in head-on collisions, fluid from the less massive parent
experiences less shock heating than fluid with the same initial
pressure from the more massive parent, (3) in off-axis collisions
with multiple periastron passages before merger, fluid from the
less massive parent experiences more shock heating than fluid
with the same initial pressure from the more massive parent, and
(4) the amount of shock heating for each parent clearly must be
the same if the two parent stars are identical.

\begin{figure}
\plotone{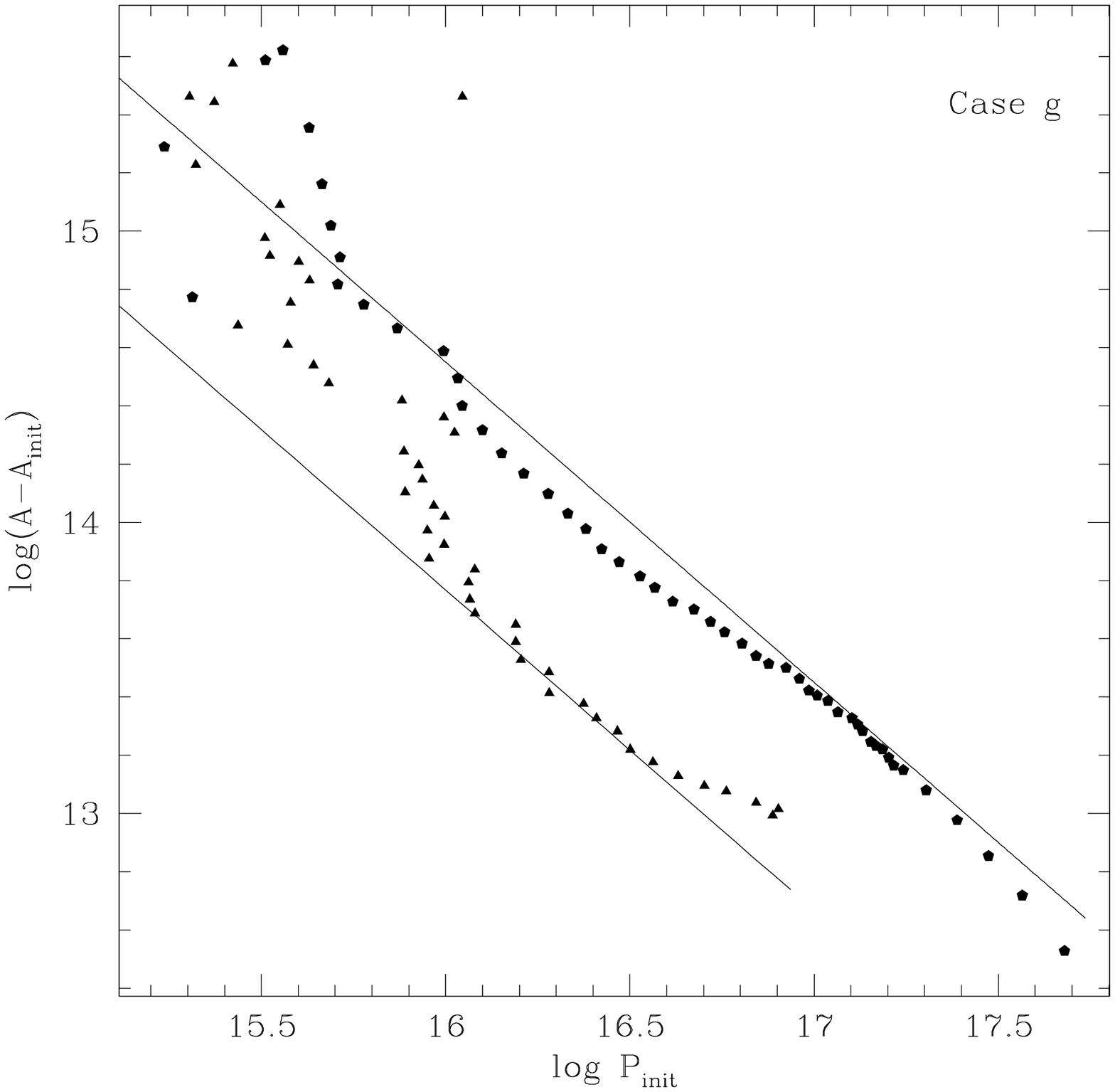} \caption{The change in entropic
variable $A$ as a function of initial pressure $P_{init}$ on a
log plot for the SPH remnant in Case~g (head-on collision of
$M_1=0.8 M_\odot$ and $M_2=0.4 M_\odot$ realistically modelled
parent stars).  Circles refer to fluid from parent star~1 which
has reached equilibrium by the end of the simulation and which
has been binned by enclosed mass fraction; triangles refer to the
corresponding fluid from parent star~2. The lines are fits to the
data with slopes of $-1.1$ and intercepts $b_i$ which differ by
$2.6\log(M_1/M_2)$ [see eqs.\ (\ref{delA}) and (\ref{b2})]. Units
are cgs. \label{glogp} }
\end{figure}

We find that when $\log(A-A_{init})$ is plotted versus
$\log(P_{init})$, the resulting curve for each parent star was
fairly linear (see Fig.\ \ref{glogp}) with a slope of
approximately $c_3=-1.1$ throughout most of the remnant in the
$\sim 20$ simulations we examined:
\begin{equation}
\log (A-A_{init}) = b_i(r_p)-c_3\log P_{init}, ~~~ i=1,2
\label{delA}.
\end{equation}
Here the intercept $b_i(r_p)$ is a function of the periastron
separation $r_p$ for the initial parabolic trajectory as well as
the masses $M_1$ and $M_2$ of the parent stars. Larger values of
$b_i$ correspond to larger amounts of shock heating in star $i$,
where the index $i=1$ for the more massive parent and $i=2$ for
the less massive parent ($M_2<M_1$). For simplicity of notation,
we have suppressed the index $i$ on the $A$, $A_{init}$ and
$P_{init}$ in equation (\ref{delA}).

The SPH data suggest that the intercepts $b_i(r_p)$ can be fit
according to the relations.
\begin{eqnarray}
b_1(r_p) & = & b_1(0) - c_4 {r_p \over R_1+R_2} \log(M_1/M_2) \label{b1}\\
b_2(r_p) & = & b_1(0) + \left(c_5 {r_p \over R_1+R_2} -c_6\right)
\log(M_1/M_2) \label{b2}
\end{eqnarray}
where $c_4=c_5=7$, $c_6=2.6$, and $b_1(0)$ is the intercept for a
head-on collision ($r_p=0$) between the two parent stars under
consideration.  Our method for determining $b_1(0)$ is discussed
in the next paragraph.  Clearly, expressions such as equations
(\ref{delA}), (\ref{b1}) and (\ref{b2}) are rather crude
approximations which lump together complicated effects from the
various stages of the fluid dynamics.  Note, however, that these
expressions do necessarily imply the desirable qualitative
features discussed above: (1) fluid with large initial pressure
$P_{init}$ (the fluid shielded by the outer layers of the star)
is shock heated less, (2) $b_2(0)<b_1(0)$, so that the smaller
star is shock heated less in head-on collisions, (3) $b_1(r_p)$
increases with $r_p$ while $b_2(r_p)$ decreases with $r_p$, so
that for sufficiently large $r_p$ we have $b_2(r_p)>b_1(r_p)$ and
the less massive star is shock heated more, and (4)
$b_1(r_p)=b_2(r_p)$ whenever $M_1=M_2$, so that identical parent
stars always experience the same level of shock heating.

Although equations (\ref{delA}), (\ref{b1}) and (\ref{b2})
describe how to distribute the shock heating, the overall
strength of the shock heating hinges on the value chosen for
$b_1(0)$.  To determine $b_1(0)$, we consider the head-on
collision between the parent stars under consideration and
exploit conservation of energy: more specifically, we choose the
value of $b_1(0)$ which ensures that the initial energy of the
system equals the final energy during a head-on collision. Since
we are considering parabolic collisions, the orbital energy is
zero and the initial energy is simply $E_{tot}=E_1+E_2$, the sum
of the energies for each of the parent stars.  The final energy
of the system includes any energy associated with ejecta and the
center of mass motion of the remnant, in addition to the energy
$E_r$ of the remnant in its own center of mass frame. Values of
$E_1$, $E_2$ and $E_r$ are the sum of the internal and
self-gravitational energies calculated while integrating the
equation of hydrostatic equilibrium. Since the energy $E_r$
depends on the structure of the remnant, it is therefore a
function of a shock heating parameter $b_1(0)$ (see \S3.3
%
%
for the details of how the remnant's structure is
determined).

The difference between the initial energy $E_{tot}$ of the system
and the energy $E_r$ of the remnant should be close to zero, but
it differs by an amount proportional to the total mass of the
ejecta:
\begin{equation}
E_{tot}=E_r-c_8 f_L E_{tot}, \label{Etot}
\end{equation}
where the coefficient $c_8$ is order unity and $f_L$ is the
fraction of mass lost during the collision (see \S3.1).
%
%
We use a value of $c_8=2.5$, which is consistent with all the
available SPH data. In equation (\ref{Etot}), the left hand side
is the initial energy of the system, and the right hand side is
its final energy.  The second term on the right hand side
accounts for the energy associated with these ejecta and with any
center of mass motion of the remnant (note this term is positive
since $E_{tot}<0$). In practice, we iterate over $b_1(0)$ until
equation (\ref{Etot}) is solved.  Equation (\ref{Etot}) needs to
be solved only once for each pair of parent star masses $M_1$ and
$M_2$: once $b_1(0)$ is known, we can model shock heating in a
collision with any periastron separation $r_p$ by first
calculating $b_1(r_p)$ and $b_2(r_p)$ from equations (\ref{b1})
and (\ref{b2}) and by then using these values in equation
(\ref{delA}).

\subsection{Merging and Fluid Mixing}\label{mixing}

As with any star in stable dynamical equilibrium, the remnant
will have an $A$ profile that increases outward.  In our model,
fluid elements with a particular $A$ value in both parent stars
will mix to become fluid in the remnant with the same value of
the entropic variable. Furthermore, if the fluid in the core of
one parent star has a lower $A$ value than any of the fluid in
the other parent star, the former's core must become the core of
the remnant, since the latter cannot contribute at such low
entropies.  When merging the fluid in the two parent stars to
form the remnant, we use the post-shock entropic variable $A$, as
determined from equation (\ref{delA}).

Within the merger remnant, the mass $m_r$ enclosed within a
surface of constant $A$ must equal the sum of the corresponding
enclosed masses in the parents:
\begin{equation}
\left. m_r\right|_{A_r=A} = \left. m_1\right|_{A_1=A} + \left.
m_2\right|_{A_2=A}.
\end{equation}
It immediately follows that the derivative of the mass in the
remnant with respect to $A$ equals the sum of the corresponding
derivatives in the parents:
d$m_r/$d$A_r=$d$m_1/$d$A_1+$d$m_2/$d$A_2$, or $dA_r/dm_r =
[(dA_1/dm_1)^{-1}+(dA_2/dm_2)^{-1}]^{-1}$.  We calculate these
derivatives using simple finite differencing.  Consequently, if we
break our parent stars and merger remnant into mass shells, then
two adjacent shells in the remnant which have enclosed masses
which differ by $\Delta m_r$ will have entropic variables which
differ by
\begin{equation}
\Delta A_r = {\Delta m_r \over \left({d A_1\over
dm_1}\right)^{-1} + \left({dA_2\over dm_2}\right)^{-1}}.
\label{DeltaAr}
\end{equation}
The value of $A$ at a particular mass shell in the remnant is then
determined by adding $\Delta A_r$ to the value of $A$ in the
previous mass shell.

In the case of the (non-rotating) remnants formed by head-on
collisions, knowledge of the $A$ profile is enough to uniquely
determine the pressure $P$, density $\rho$, and radius $r$
profiles.  While forcing the $A$ profile to remain as was
determined from sorting the shocked fluid, we integrate
numerically the equation of hydrostatic equilibrium with d$m=4\pi
r^2\rho$d$r$ to determine the $\rho$ and $P$ profiles [which are
related through $\rho=(A/P)^{3/5}$]. This integration is an
iterative process, as we must initially guess at the central
pressure.  Our boundary condition is that the pressure must be
zero when the enclosed mass equals the desired remnant mass
$M_r=(1-f_L)(M_1+M_2)$.  During this numerical integration we also
determine the remnant's total energy $E_r$ and check that the
virial theorem is satisfied to high accuracy. The total remnant
energy $E_r$ appears in equation (\ref{Etot}), and if this
equation is not satisfied to the desired level of accuracy, we
adjust our value of $b_1(0)$ accordingly and redo the shocking
and merging process.

Once the $A$ profile of the remnant has been determined, we focus
our attention on its chemical abundance profiles.  Not all fluid
with same initial value of $A_{init}$ is shock heated by the same
amount during a collision, since, for example, fluid on the
leading edge of a parent star is typically heated more violently
than fluid on the trailing edge of the parent.  Consequently,
fluid from a range of initial shells in the parents can
contribute to a single shell in the remnant.  To model this
effect, we first mix each parent star by spreading out its
chemicals over neighboring mass shells, using a Gaussian-like
distribution which depends on the difference in enclosed mass
between shells.  Let $X_i$ be the chemical mass fraction of some
species $X$ in a particular shell $i$, and let the superscripts
``pre'' and ``post'' indicate pre- and post-mixing values,
respectively.  Then
\begin{eqnarray}
X_k^{post} & = & \sum_i{X_i^{pre} g_{ik} \Delta m_i \over \sum_j g_{ji} \Delta m_j}, \label{Xkpost} \\
g_{ik} & = & \exp\left[-{\alpha \over M^2}(m_i-m_k)^2\right]\nonumber\\
 && +\exp\left[-{\alpha \over M^2}(m_i+m_k)^2\right]+\exp\left[-{\alpha \over M^2}(m_i+m_k-2M)^2\right], \label{gik}\\
\alpha & = & c_8 \left( M \overline{{d\ln A \over dm}}\right)^2,
\label{alpha}
\end{eqnarray}
where $\Delta m_i$ is the mass of shell $i$, $m_i$ is the mass
enclosed by shell $i$, $M$ is the total mass of the parent star,
and $c_8$ is a dimensionless coefficient which we choose to be
1.5. We have suppressed an additional index in equations
(\ref{Xkpost}) through (\ref{alpha}) which would label the parent
star. The summand in equation (\ref{Xkpost}) is the contribution
from shell $i$ to shell $k$.  The second term in the distribution
function equation (\ref{gik}) is important only for mass shells
near the center of the parent, while the third term becomes
important only for mass shells near the surface; these two
correction terms guarantee that an initially chemically
homogeneous star remains chemically homogeneous during this
mixing process ($X_k^{post}=X_k^{pre}=$constant, for any shell
$k$).  The bar in equation (\ref{alpha}) represents an average
over the parent star, and the dependence of $\alpha$ on the
average $d\ln A/dm$ ensures that stars with steep entropy
gradients are more difficult to mix (see Table 4 of Lombardi et
al.\ 1996).

Consider a fluid layer of mass $dm_r$ in the merger remnant with
an entropic variable $A$ which ranges from $A_r$ to $A_r+dA_r$.
The fraction of that fluid $dm_i/dm_r$ which originated in parent
star $i$ can be calculated as $(dA_r/dm_r)/(dA_i/dm_i)$.
Therefore, the composition of this fluid element can be
determined by the weighted average
\begin{equation}
X_r = X^{post,1} {dA_r/dm_r \over dA_1/dm_1} + X^{post,2}
{dA_r/dm_r \over dA_2/dm_2}, \label{xr}
\end{equation}
where all derivatives are evaluated at $A_r$, the value of the
entropic variable under consideration.  Equation (\ref{xr})
allows us to determine the final composition profile of any
merger remnant simply from the $A$ profiles of the parent stars
and merger remnant, as well as the (post-mixed) composition
profile of each parent as given by equation (\ref{Xkpost}).

\subsection{Angular Momentum Distribution}

To estimate the total angular momentum $J_r$ of the remnant in
its center of mass frame, we use angular momentum conservation in
the same way that energy conservation was used in
\S\ref{shock}~~In particular, the total angular momentum in the
system is given by
\begin{equation}
J_{tot}=M_1M_2\left({2 G r_p\over M_1+M_2}\right)^{1/2},
\end{equation}
where $G$ is Newton's gravitational constant, and this angular
momentum must equal $J_r$ plus a contribution due to mass loss
[cf.\ eq.\ (\ref{Etot})]:
\begin{equation}
J_{tot}=J_r+c_{10} f_L J_{tot}. \label{Jtot}
\end{equation}
The SPH simulations demonstrate that $J_{tot}$ is always slightly
larger than $J_r$, and the choice $c_{10}=2.5$ does a good job of
replicating the SPH results.  Equation (\ref{Jtot}) can be solved
for $J_r$, and the results are compared with those of SPH
simulations for a few typical cases in Table \ref{tbl-3}.

The structure of the rotating remnants formed in off-axis
collisions depends on the distribution of this angular momentum
within the remnant.  The goal here is to simplify this
complicated distribution (see Fig.\ \ref{all}) into an average
one-dimensional profile that can be used in YREC.  The specific
angular momentum $j$ profile for the SPH remnants increases
outward and is typically concave upward throughout most of the
remnant when averaged over isodensity surfaces and plotted against
enclosed mass. Once an approximate analytic {\it form} for this
average $j$ profile (which is only weakly dependent on the
periastron separation $r_p$) is specified, the profile can be
normalized simply by requiring
\begin{equation}
J_r=\int_0^{M_r} j(m) dm, \label{Jr}
\end{equation}
where $m$ corresponds to the mass enclosed within a constant
density surface and $J_r$ is determined from equation
(\ref{Jtot}).  We find that, unlike a simple power-law
dependence, the relation $j(m)\propto (\exp(c_9m/M_r)-1)$ (with
$c_9\approx 2.3$) is able to reproduce the qualitative features
of the angular momentum profile. Using equation (\ref{Jr}) to
find the proportionality coefficient, one obtains
\begin{equation}
j(m)={c_9\over \exp(c_9)-1-c_9}{J_r \over
M_r}\left(\exp\left(c_9{m\over M_r}\right)-1\right). \label{jofm}
\end{equation}
Other forms for $j(m)$ could also be used, being normalized
through equation (\ref{Jr}).

\begin{table}
\caption{Total Angular Momentum \label{tbl-3}}
\begin{tabular}{cccc}
\\
Case & $J_{tot}$ [g cm$^2$/s]  & $J_{r,SPH}$ [g cm$^2$/s] & $J_r$ [g cm$^2$/s]\\
\tableline
E & 2.1$\times10^{51}$& 2.0$\times10^{51}$& 2.0$\times10^{51}$\\
I & 2.0$\times10^{51}$& 2.0$\times10^{51}$& 2.0$\times10^{51}$\\
e & 2.1$\times10^{51}$& 2.0$\times10^{51}$& 2.0$\times10^{51}$\\
f & 3.0$\times10^{51}$& 2.8$\times10^{51}$& 2.9$\times10^{51}$\\
k & 1.4$\times10^{51}$& 1.4$\times10^{51}$& 1.4$\times10^{51}$
\end{tabular}
\end{table}

\subsection{Comparisons with SPH Results}

To test further the accuracy of our simple models, we compare
them to the results of SPH simulations.  For the two collisions
presented Sills \& Lombardi 1997, the realistic parent star
models were created using YREC.  The first collision, Case~a, is
between two (turnoff) $0.8\,M_\odot$ stars, and the second,
Case~g, is between a $0.8M_\odot$ star and a $0.4M_\odot$ star
(see Tables \ref{tbl-1} and \ref{tbl-2} for more details).

For Case~a, the total energy of the system is
$E_{tot}=-5.23\times 10^{48}$ erg, and the total energy of the
remnant is $E_r=-6.33\times 10^{48}$ erg [see eq.\
(\ref{Etot})].  For Case~g, these energies are $-3.35\times
10^{48}$ erg and $-3.82\times 10^{48}$ erg.  For Case~a,
$\alpha=308$ for the parents; for Case~g, $\alpha_1=310$ for the
$0.8M_\odot$ parent and $\alpha_2=92$ for the $0.4M_\odot$ parent.

\begin{figure}
\plotone{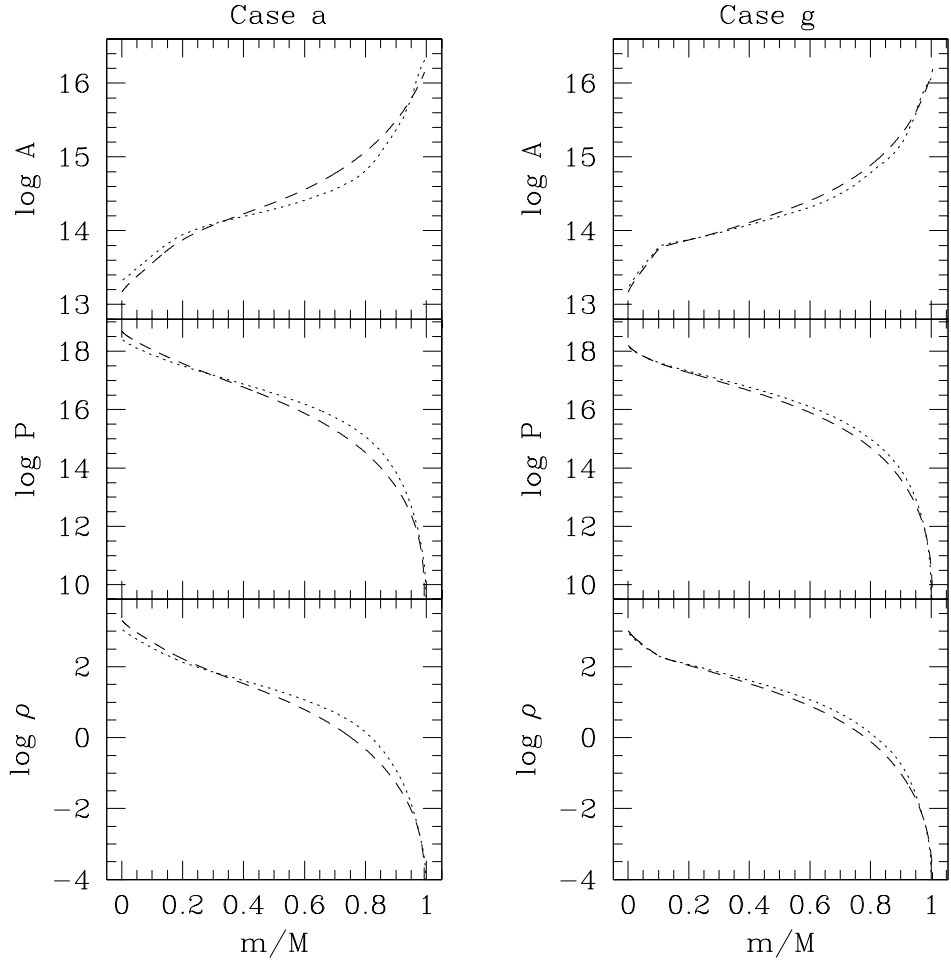} \caption{Thermodynamic profiles of
$A$, pressure $P$, and density $\rho$ as a function of enclosed
mass fraction $m/M$ for the remnants of Cases a and g, where $M$
is the total bound mass of the remnant.  The dotted line refers
to the remnant resulting from a 3D SPH simulation, and the dashed
line refers to the remnant generated by the method of this paper.
All units are cgs. \label{thermag} }
\end{figure}

\begin{figure}
\plotone{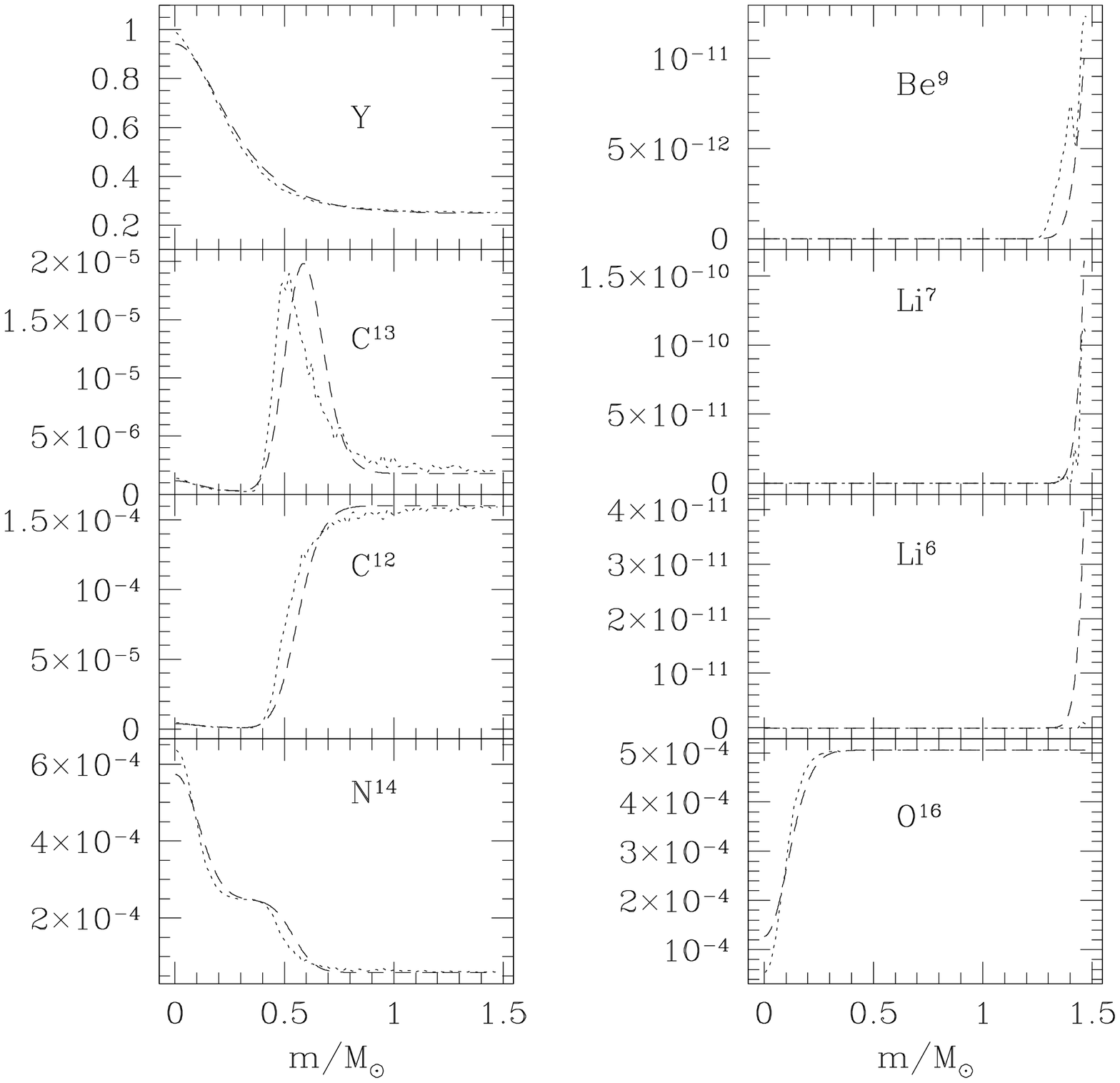} \caption{Fractional chemical
abundance (by mass) as a function of enclosed mass fraction $m/M$
for the Case~a remnant.  Line types are as in Fig.\
\ref{thermag}. \label{chem8a} }
\end{figure}

Thermodynamic (Fig.\ \ref{thermag}) and chemical (Fig.\
\ref{chem8a}) profiles show that our remnant models are quite
accurate.  In Case~g, our remnant displays the kink in the $A$
profile near $m/M=0.1$ (see Fig.\ \ref{thermag}), inside of which
fluid originates solely from the $0.8M_\odot$ star.  Our models
also reproduce the chemical profiles of the SPH remnant very well:
the peak values in the chemical abundances are accurate to within
roughly ten percent, and the shapes of these profiles, though
sometimes peculiar, are followed closely.  Helium distribution is
particularly important to model well since it will determine the
MS lifetime of the remnant.  The central values of the fractional
helium abundance $Y$ given by our model differ from the SPH
result by only about 0.05.

\begin{figure}
\plotone{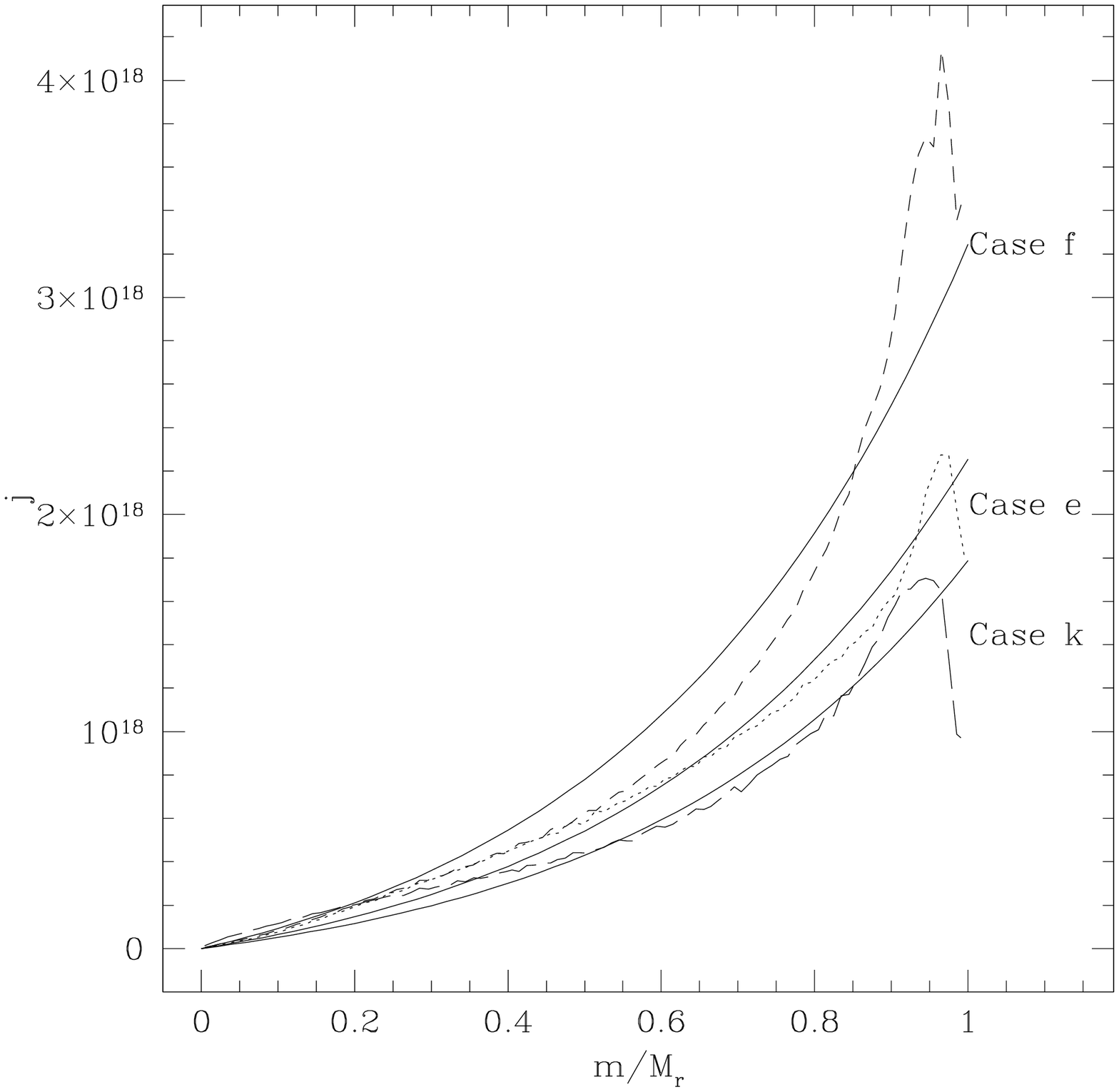} \caption{SPH specific angular
momentum profiles averaged over surfaces of constant density
(dotted or dashed curves) compared with the approximate profiles
(solid curves) generated from equation (\ref{jofm}). Profiles are
plotted against the enclosed mass fraction $m/M_r$, with values
of the remnant mass $M_r$ being given in Table \ref{tbl-2}.
\label{jreal2.3} }
\end{figure}

Similar levels of agreement are found between our models of
rotating remnants and their corresponding SPH models.  Specific
angular momentum profiles, averaged over surfaces of constant
density, are compared in Figure \ref{jreal2.3} for cases e, f and
k.  Note that the hump in the SPH $j$ profile in the outer few
percent of the remnant is due to having to terminate the
simulation before all of the gravitationally bound fluid has
fallen back to the merger remnants: this feature is still
diminishing gradually during the final stage of the SPH
calculation and hence we do not attempt to model it.

It is interesting to note that there is very little lithium in our
remnants.  Lithium is burned during stellar evolution except at
very low temperatures ($\la 2\times 10^7$ K), and therefore can
be used as an indicator of mixing.  If a star has a deep enough
surface convective layer, there will be essentially no lithium,
because the convection mixes any lithium from the outer layers
into the hot interior where it is burned.  A small amount of
lithium in our $0.8 M_\odot$ parent star does exist in the outer
few percent of its mass and consequently becomes part of the
ejecta during the collision. Although near the remnant's surface
our method yields a large {\it fractional\/} error in lithium
abundance (see Fig.\ \ref{chem8a}), this is simply because the
overall abundance is so close to zero.  For example, the
predicted surface fractional Li$^6$ abundance of $4 \times
10^{-11}$ for our Case~a remnant is an overestimate, but is
roughly 30 times smaller than the surface fractional abundance in
the $0.8 M_\odot$ parent.  Except for in the extreme case of
grazing collisions (in which mass loss would be exceedingly
small), collisional blue stragglers should be severely depleted
in lithium, a prediction that can be tested with appropriate
observations (see Shetrone \& Sandquist 2000 for a recent attempt
at detecting lithium in blue stragglers).

Since our method takes considerably less than a minute to generate
a model on a typical workstation, versus hundreds or thousands of
hours for a hydrodynamic simulation to run on a supercomputer, we
are able to explore the results of collisions in a drastically
shorter amount of time. Our approach is easily generalized to work
for more than two parent stars, by colliding two stars first and
then colliding the remnant with a third parent star.  Such
algorithms will make it possible to incorporate the effects of
collisions in simulations of globular clusters as a whole.

\acknowledgments We would like to thank Joshua Faber, Jessica
Sawyer, Alison Sills, and Aaron Warren for their many
contributions to this work.  J.C.L.\ acknowledges support from the
Keck Northeast Astronomy Consortium, from a grant from the
Research Corporation, and from NSF Grant AST-0071165. F.A.R.\
acknowledges support from NSF Grants AST-9618116 and PHY-0070918,
NASA ATP Grant NAG5-8460, and a Sloan Research Fellowship. This
work was also partially supported by the National Computational
Science Alliance under Grant AST980014N and utilized the NCSA
SGI/Cray Origin2000 parallel supercomputer.

\end{document}